\documentstyle[twoside,fleqn,espcrc2]{article}

\newcommand{\AmS}{{\protect\the\textfont2
  A\kern-.1667em\lower.5ex\hbox{M}\kern-.125emS}}
\title{On reflection positive formulation of chiral gauge theories
       on a lattice}

\author{Sergei V. Zenkin\address{Institute for Nuclear Research of
        Russian Academy of Sciences, 60th October Anniversary Prospect
        7a, 117312 Moscow, Russia}}

\begin{document}

\begin{abstract}
A formulation of chiral gauge theories on a lattice which is both
reflection positive and gauge invariant is discussed.
\end{abstract}

\maketitle

\section{INTRODUCTION}

In constructing continuum quantum gauge theories from the lattice
ones the properties of reflection positivity and gauge invariance
of the latter play a fundamental role (see, e.g., \cite{Seiler} and
references therein). Reflection positivity allows one to construct
the Hilbert space of states with positive definite metric and ensures
the canonical quantum mechanical interpretation of the theories. This
guarantees their unitarity, that is particularly important for gauge
theories. Gauge invariance controls a space of relevant parameters of
the theory and facilitates an efficient tuning to the continuum limit.

Vector lattice gauge theories in the Wilson formulation are naturally
reflection positive and gauge invariant, while for chiral ones it is
a long standing problem to satisfy these properties. The only
exception is mirror fermion model \cite{Montvay}. It, however, is
actually chiral only provided the chiral symmetry is spontaneously
broken. Here we consider another formulation of chiral gauge theories
on a lattice which is both reflection positive and gauge invariant
\cite{ZZ}.

\section{CONVENTIONS}

We consider a hypercubic D-dimensional (D is even) lattice $\Lambda$
with sites numbered by $n = (n_{0}, \ldots, n_{D-1})$, $-N/2+1 \le
n_{\mu} \le N/2$, $N$ is even, with a lattice spacing $a$; $\hat{\mu}$
is the unit vector along a lattice link in the positive
$\mu$-direction. We shall define a theory on a torus $T_{D}$ which is
obtained by addition of links connecting each pair of sites with
$n_{\mu} = N/2$ and $n_{\mu} = -N/2+1$.

Let the gauge group $G$ be unitary. Dynamical variables of the theory
are the fermion $2^{D/2}$-component (Grassmannian) fields $\psi_{n}$,
$\overline{\psi}_{n}$, defined on lattice sites, and gauge variables
$U_{n, n + \hat{\mu}}
\in G$, $U_{n, n - \hat{\mu}} = U^{\dag}_{n -\hat{\mu}, n}$,
defined on lattice links. Conventionally $U_{n, n +
\hat{\mu}} = \exp[i\,g\, a\, A_{\mu}(n +
\hat{\mu}/2)]$, where $A$ is a gauge field which belongs to the
gauge group algebra and $g$ is gauge coupling. We shall use the
representation for $U$ introduced in ref.
\cite{BFS} in terms of "half gauge" variables $W_{(n, \pm
\hat{\mu})} \in G$ associated with each pair $(n, \pm
\hat{\mu})$:
\begin{equation}
U_{n, n + \hat{\mu}} = W_{(n, \hat{\mu})} \; W^{\dag}_{(n + \hat{\mu},
-\hat{\mu})}.
\end{equation}
For $W$ there is no representation in terms of a local field which
transforms as an irreducible representation of the group of rotation
of the Euclidean space. However to allow for a perturbative
consideration we introduce variables $z_{\pm \hat{\mu}}(n)$, so that
$W_{(n, \pm \hat{\mu})} = \exp[i\,g\, a\, z_{\pm \hat{\mu}}(n)]$.

For a simplicity we consider right-handed fermions being
singlets under the gauge group, so the gauge transformations are
defined as follows:
\begin{eqnarray}
\lefteqn{\psi_{n} \rightarrow (h_{n} P_{L} + P_{R}) \psi_{n},}
\nonumber \\
\lefteqn{\overline{\psi}_{n} \rightarrow \overline{\psi}_{n}
(h_{n}^{\dag} P_{R} + P_{L}),}  \\
\lefteqn{W_{(n, \pm \hat{\mu})} \rightarrow
h_{n} \; W_{(n, \pm \hat{\mu})},}
\end{eqnarray}
where $h_{n} \in G$, $P_{L,R} = (1 \pm \gamma_{D+1})/2$ are chiral
projecting operators. For general case of non-singlet $P_{R} \psi$
see \cite{ZZ}.

A theory with an action $A[\psi, \overline{\psi}, W]$ is defined by
the functional integrals
\begin{eqnarray}
\lefteqn{Z^{-1}\;\int \prod_{n \in \Lambda} d \psi_{n} d \overline{\psi}_{n}
\prod_{n
\in \Lambda, \mu} d W_{(n, \hat{\mu})} d W_{(n, -\hat{\mu})}}\nonumber \\
\lefteqn{\cdot O[\psi, \overline{\psi}, W] \; e^{-\displaystyle A},}
\end{eqnarray}
where $Z$ is the partition function of the theory, $d W_{(n, \pm
\hat{\mu})}$ is the Haar measure, and $O$ is some functionals of the
dynamical variables.

Let $\Lambda_{\pm}$ denote the equal parts of the lattice with $n_{0}
> 0$ and $n_{0} < 0$, respectively, and let $r$ be such a reflection,
that $r \Lambda_{\pm} = \Lambda_{\mp}$. So, the reflection does not
change $n_{\mu}$ and $\hat{\mu}$ for $\mu \neq 0$, while $r n_{0} = -
n_{0} + 1$, $r \hat{0} = - \hat{0}$.

Given reflection $r$, an antilinear operator $\theta$ is defined as
\begin{eqnarray}
\lefteqn{\theta[\overline{\psi}_{m} \: \Gamma \: W_{(m, \pm
\hat{\mu})} \cdots \psi_{n}]}
\nonumber \\
\lefteqn{= \overline{\psi}_{r n} \gamma_{0} \cdots W^{\dag}_{r (m, \pm
\hat{\mu})} \: \Gamma^{\dag} \:
\gamma_{0} \psi_{r m},}
\end{eqnarray}
where $\Gamma$ is a matrix.

A theory is called reflection positive if for each functional $O$ of
the form $F \theta [F]$, where $F = F[\psi, \overline{\psi}, W]$ is
defined on $\Lambda_{+}$, the integral (4) is non-negative.  The
sufficient condition for a theory with an action $A$ to be reflection
positive is existence of such functionals $B[\psi,
\overline{\psi}, W]$ and $C_{i}[\psi, \overline{\psi}, W]$ defined on
$\Lambda_{+}$ that $A$ can be represented in the form \cite{BFS}
\begin{equation}
- A = B + \theta [B] + \sum_{i} C_{i} \theta [C_{i}].
\end{equation}

\section{CONSTRUCTING THE THEORY}

We proceed from the Wilson action for free massless fermions:
\begin{eqnarray}
\lefteqn{A = a^{D} \sum_{n \in \Lambda, \mu}
\overline{\psi}_{n} \biggl[
\gamma_{\mu} \frac{1}{2a} (\psi_{n +
\hat{\mu}} - \psi_{n - \hat{\mu}})} \nonumber \\
\lefteqn{- \frac{1}{2a} (\psi_{n + \hat{\mu}} + \psi_{n -
\hat{\mu}} - 2 \psi_{n})\biggr],}
\end{eqnarray}
where $\psi_{(\ldots, N/2 + 1, \ldots)} = - \psi_{(\ldots, -N/2 + 1,
\ldots)}$, $\psi_{(\ldots, -N/2, \ldots)} = - \psi_{(\ldots, N/2,
\ldots)}$. This is the simplest form of the lattice fermion action
which is determined by the finite dimension approximation of
functional integrals for canonical Hamiltonian (Grassmannian)
dynamics and satisfies condition (6) of reflection positivity
\cite{Zen2}.

Action (8) is not invariant under the global transformation of the
form (2). Therefore, we seek the gauge action in the form
\begin{eqnarray}
\lefteqn{A=a^{D} \sum_{n \in \Lambda, \mu} \overline{\psi}_{n} \biggl[
\gamma_{\mu}\frac{1}{2a} \biggl((P_{L}U_{n, n+\hat{\mu}}
+P_{R}) \psi_{n +\hat{\mu}}} \nonumber \\
\lefteqn{-(P_{L}U_{n, n-\hat{\mu}} + P_{R})
\psi_{n - \hat{\mu}}\biggr)} \nonumber \\
\lefteqn{-\frac{1}{2a}\biggl(
(P_{L}X^{L}_{n, n+\hat{\mu}}+ P_{R}X^{R}_{n, n+\hat{\mu}})\psi_{n +
\hat{\mu}}} \nonumber \\
\lefteqn{+ (P_{L}X^{L}_{n, n-\hat{\mu}}
+P_{R}X^{R}_{n, n-\hat{\mu}})
\psi_{n - \hat{\mu}}} \nonumber \\
\lefteqn{- 2 (P_{L}Y^{L}_{(n, \hat{\mu})} + P_{R}Y^{R}_{(n, \hat{\mu})})
\psi_{n}\biggr) \biggr],}
\end{eqnarray}
where $X$ and $Y$ are some functions of $W$.

Let us require action (8) to be invariant under rotations of the
lattice by $\pi/2$, and to satisfy the following conditions: (i)
condition (6) of reflection positivity; (ii) gauge invariance; (iii)
in the ungauged limit $W =1$ it takes the form of eq. (7). Then, from
(i) we find
\begin{eqnarray}
\lefteqn{X^{L}_{n, n + \hat{\mu}} = W^{\dag}_{(n + \hat{\mu}, -\hat{\mu})},}
\nonumber \\
\lefteqn{X^{R}_{n, n + \hat{\mu}} = W_{(n, \hat{\mu})},}
\end{eqnarray}
while from (ii), (iii) one has
\begin{eqnarray}
\lefteqn{Y^{L}_{(n,\hat{\mu})} = \frac{1}{2}(W^{\dag}_{(n, \hat{\mu})} +
W^{\dag}_{(n, -\hat{\mu})}),}
\nonumber \\
\lefteqn{Y^{R}_{(n,\hat{\mu})} =
\frac{1}{2}( W_{(n, \hat{\mu})} + W_{(n, - \hat{\mu})}).}
\end{eqnarray}
So these requirements determine the action uniquely.

One can rewrite the action and the measure in the functional
integrals in terms of $U_{n, n \pm \hat{\mu}}$ and, say,  $W_{(n,
\hat{\mu})}$. Then our theory is determined by functional integrals
of the form
\begin{eqnarray}
\lefteqn{Z^{-1} \int \prod_{n \in \Lambda} d \psi_{n} d \overline{\psi}_{n}
\prod_{n \in \Lambda, \mu} d U_{n, n + \hat{\mu}}
d W_{(n, \hat{\mu})}} \nonumber \\
\lefteqn{\cdot O[\psi, \overline{\psi}, U, W]
e^{-\displaystyle A_{gauge} + A[\psi, \overline{\psi}, U, W]},}
\end{eqnarray}
where $A_{gauge}$ is a reflection positive action for gauge variables
and $O$ is a gauge invariant functional of the dynamical variables.

\section{DISCUSSION}

Owing to reflection positivity this theory is unitary, but, in
general, this holds for full Hilbert space including all gauge
variables: either $W_{n, \hat{\mu}}$ and $W_{(n, -\hat{\mu})}$, or
$U_{n, n + \hat{\mu}}$ and $W_{(n, \hat{\mu})}$. An argument for that
the theory may not be unitary in the subspace of the conventional
variables $\psi$, $\overline{\psi}$, $U$ is that explicit integrating
over $W_{(n, \hat{\mu})}$ in (11) for operators $O$ independent of
such variables leads to a theory whose action does not satisfy
condition (6) of reflection positivity. Therefore we must require that
unpaired variables $W$ decouple. The price to be paid for this is
the main question to this approach.

If $A_{gauge}$ in (11) is the Wilson plaquette action, formal limit
of the full action at $a \rightarrow 0$ coincides with the action of
the continuum chiral gauge theory with dynamical variables $\psi$,
$\overline{\psi}$, and $A$ (target theory \cite{Roma1}):
\begin{equation}
A = \int d^{D} \! x \, [\frac{1}{4} F_{\mu \nu}F_{\mu \nu} +
\overline{\psi}\gamma_{\mu} (D_{\mu} P_{L} + \partial_{\mu} P_{R})
\psi ],
\end{equation}
where $D_{\mu}$ is covariant derivative. However decoupling $W$ at
the classical level does not guarantee their decoupling in the quantum
theory. This is true, however, of right-handed fermions, because
action (8) has the shift symmetry $\psi_{n} \rightarrow \psi_{n} +
P_{R} \epsilon$, $\overline{\psi}_{n}
\rightarrow \overline{\psi}_{n} + \overline{\epsilon} P_{L}$, that
guarantees their decoupling in the continuum limit \cite{GP}.

To get some idea of what happens to unpaired $W$ we consider in this
formulation the chiral Schwinger model \cite{JR}, whose perturbative
solution (at least in the topologically trivial sector) is known to be
exact.

\subsection{Two-dimensional example}

Let us consider the continuum limit of the effective action
\begin{eqnarray}
\lefteqn{W[U, W] = - \ln \int \prod_{n \in \Lambda} d \psi_{n} d
\overline{\psi}_{n}} \nonumber \\
\lefteqn{\cdot e^{-\displaystyle A[\psi, \overline{\psi}, U, W]}.}
\end{eqnarray}

Then, for sufficiently smooth $A$ and $z$, we find
\begin{eqnarray}
\lefteqn{W[A, z]=\frac{1}{2} \int \frac{d^{2} q}{(2 \pi)^2}
\sum_{\mu, \nu} \biggl[ A_{\mu}(-q) \; [ \delta_{\mu \nu} q^{2} -
q_{\mu} q_{\nu}} \nonumber \\
\lefteqn{+ \Pi^{AA}_{\mu \nu}(q) ] \; A_{\nu}(q)
+ \bar{A}_{\mu}(-q)\; \Pi^{\bar{A} \bar{A}}_{\mu \nu}\;
\bar{A}_{\nu}(q)}
\nonumber \\
\lefteqn{+ \bar{A}_{\mu}(-q) \;\Pi^{\bar{A} z}_{\mu \nu}\;
z_{\hat{\nu}}(q)
+ z_{\hat{\mu}}(-q) \;\Pi^{z \bar{A}}_{\mu \nu}\; \bar{A}_{\nu}(q)}
\nonumber \\
\lefteqn{+ z_{\hat{\mu}}(-q) \;\Pi^{z z}_{\mu \nu} \;z_{\hat{\nu}}(q)}
\nonumber \\
\lefteqn{+ i \bar{A}_{\mu}(-q)\; K^{\bar{A} \bar{A}}_{\mu \nu}(q)
\;\bar{A}_{\nu}(q)
\biggr].}
\end{eqnarray}
Here $\bar{A}_{\mu}(q)$ is $a \rightarrow 0$ limit of the gauge
invariant combination ${A}_{\mu}(q) + 2i \sin (\frac{1}{2}q_{\mu}a) \;
z_{\hat{\mu}}(q)$,
\begin{eqnarray}
\lefteqn{\Pi^{AA}_{\mu \nu}(q) = \frac{g^{2}}{2 \pi} (\delta_{\mu \nu} -
\frac{q_{\mu} q_{\nu}}{q^{2}}),}\nonumber \\
\lefteqn{K^{\bar{A} \bar{A}}_{\mu \nu}(q) = \frac{g^{2}}{4 \pi}
\frac{1}{q^{2}} (\epsilon_{\mu
\alpha} q_{\alpha} q_{\nu} + q_{\mu} \epsilon_{\nu \alpha} q_{\alpha}),}
\end{eqnarray}
and $\Pi^{\bar{A} \bar{A}}_{\mu \nu}$, $\Pi^{\bar{A} z}_{\mu \nu} =
\Pi^{z \bar{A}}_{\mu \nu}$, and $\Pi^{z z}_{\mu \nu}$ are some
symmetrical matrices independent of $q$. From the Ward identities and
the lattice rotation symmetry we have
\begin{equation}
\sum_{\mu} \Pi^{z \cdot}_{\mu \nu} = \sum_{\nu} \Pi^{\cdot
z}_{\mu \nu} = 0,
\end{equation}
and
\begin{equation}
\Pi_{0 0} = \Pi_{1 1}, \;
\Pi^{\bar{A} z}_{\mu \nu} = - \frac{1}{2} \Pi^{z z}_{\mu \nu}, \;
\Pi^{\bar{A} z}_{0 1} = - 2 \Pi^{\bar{A} \bar{A}}_{0 1},
\end{equation}
respectively. Numerically the elements of these matrices depend on
infrared regularization used and do not make much sense; we started
with the finite lattice which ensures such a regularization, then
$\Pi^{\bar{A}
\bar{A}}_{0 0} = 1.959(7) g^{2}/(2 \pi)$, $\Pi^{\bar{A} \bar{A}}_{0 1}
= - 0.361(6) g^{2}/(2 \pi)$.

The lessons from this example are following: The formulation ensures
decoupling the doubler fermion modes in the external gauge fields.
Gauge invariance of the effective action in the case of anomaly
fermion contents is provided by producing a Wess-Zumino term. It
involves variable $z$ instead of scalar one and reads as $(i/2)
\int_{q}\, \sum_{\mu, \nu}\,(\bar{A}_{\mu}\, K^{\bar{A} \bar{A}}_{\mu
\nu}\,\bar{A}_{\nu} - A_{\mu}\, K^{\bar{A} \bar{A}}_{\mu
\nu} \,A_{\nu})$. Variable $z$ does not decouple in quantum theory
rendering it non-invariant under continuum rotations. Therefore
additional efforts are necessary for decoupling this variable.

\subsection{Outlooks}

Let us note that in terms of $W_{n, \hat{\mu}}$ and $W_{(n,
-\hat{\mu})}$ the principal difference between vector and chiral
gauge theories is that in the vector theories variables $W$ are
always paired forming link variables $U$, while in chiral ones they
are splitted by the Wilson term. This means that vector gauge
theories have additional symmetry as compared to chiral ones. This
symmetry can be defined as invariance of the theory under
transformations:
\begin{equation}
W_{(n, \pm \hat{\mu})} \rightarrow  W_{(n,
\pm \hat{\mu})} \; g_{n \pm \hat{\mu}/2},
\end{equation}
where $g_{n \pm \hat{\mu}/2} \in G$, other variables being
non-transformed. Therefore to ensure decoupling of unpaired $W$ we can
require symmetry (18) to hold in the continuum limit.

Obviously this can be achieved by adding to the original action a set
of gauge invariant counterterms, so that the theory remains gauge
invariant under gauge transformations (2), (3) at any $a$, but
becomes invariant under both transformations (2), (3) and (18) only
in the continuum limit. As our two-dimensional example shows the
number and explicit structure of such counterterms may crucially
depend on whether the theory is anomaly one or not. Indeed, in the
case of anomaly free fermion contents, i.e. when counterpart of
$K^{\bar{A} \bar{A}}$ in (14) vanishes, symmetry (18) is restored in
the continuum limit by local counterterms of the form
\begin{equation}
\sum_{n; \mu, \nu; \eta_{1}, \eta_{2}} W^{\dag}_{(n, \eta_{1}\hat{\mu})}
Z^{\eta_{1} \eta_{2}}_{\mu \nu} W_{(n,
\eta_{2}\hat{\mu})},
\end{equation}
where $\eta_{1, 2} = \pm$. Note that these counterterms are
reflection positive. Owing to relations (16), (17) tuning only two
parameters is needed. Then in the continuum limit we come to the
Euclidean and gauge invariant in the conventional sense theory
(Jackiw's parameter \cite{JR} being $a = 1$). But to render
the anomaly theory to be invariant under (18) a non-local counterterm
is needed, as it follows from (14), (15).

Similar picture is also expected to hold in four dimensions with the
gauge fields being dynamical. A likely scenario is that in anomaly
free case the theory is determined by tuning a few relevant
parameters corresponding to gauge invariant local counterterms, while
for an anomaly theory an infinite set of counterterms, including
non-local ones, is required. An argument in favour this is that the
original action involves no couplings, except the gauge ones and in
the terms of variables $\psi$, $\overline{\psi}$, $z_{\mu}$, and
$z_{-\mu}$ the theory is renormalizable by power counting.
Therefore at small non-renormalized coupling the well-known
perturbative results should be reproduced. Then for non-abelian gauge
group the values of those relevant parameters could be determined
perturbatively due to asymptotical freedom. Certainly, if a chiral
gauge theories do exist beyond the perturbation theory and are unique
we shall come in the continuum limit to the result of Rome approach
\cite{Roma1}, \cite{Roma2}, but with gauge fixing in principle being
unnecessary and with less number of counterterms. In other words
reflection positivity and gauge invariance hopefully allows one to
project in the continuum limit enlarged Hilbert space (where the
theory is unitary at any $a$) to the physical one by gauge invariant
way not violating the unitarity. This crucially differs this
formulation from models with the Wilson-Yukawa couplings
\cite{Roma2}.

Of course, there is a lot of work to do for establishing actual
status of this formulation, first of all, perturbative calculations
in four dimensions should be done.

\section{ACKNOWLEDGMENTS}

I am grateful to J. Jers\'{a}k, H. Joos, M. L\"{u}scher, I. Montvay,
G. M\"{u}nster, D. Petcher, E. Seiler, L. Stuller, T. E. Tomboulis,
and M. Tsypin for useful discussions and suggestions. It is a
pleasure to thank J. Smit and the organizers of Lattice '92 for
financial support giving me the opportunity to attend the conference.

\end{document}